\documentclass[12pt]{article}
\usepackage{epsfig}
\usepackage{graphicx}
\usepackage{lineno}

\newcommand\pubnumber{SNSN-323-63}
\newcommand\pubdate{\today}

\def\Sussex{Department of Physics and Astronomy\\
University of Sussex, Brighton, Sussex, BN1 9QH, United Kingdom}

\def\Title#1{\begin{center} {\Large #1 } \end{center}}
\def\Author#1{\begin{center}{ \sc #1} \end{center}}
\def\Address#1{\begin{center}{ \it #1} \end{center}}

\newcommand\pubblock{\rightline{\begin{tabular}{l} \pubnumber\\
         \pubdate  \end{tabular}}}

\newenvironment{Presented}{\begin{quotation} \begin{center} 
             PRESENTED AT\end{center}\bigskip 
      \begin{center}\begin{large}}{\end{large}\end{center} \end{quotation}}

\def\beq{\begin{equation}}
\def\eeq#1{\label{#1}\end{equation}}
\def\eeqn{\end{equation}}

\def\beqa{\begin{eqnarray}}
\def\eeqa#1{\label{#1}\end{eqnarray}}
\def\eeqan{\end{eqnarray}}

\let\bar=\overbar

\def\Dslash{\not{\hbox{\kern-4pt $D$}}}
\def\dslash{\not{\hbox{\kern-2pt $\del$}}}

\def\msb{{\bar{\ssstyle M \kern -1pt S}}}

\def\Title#1{\begin{center} {\Large {\bf #1} } \end{center}}

\begin{document}
\begin{titlepage}
\pubblock

\vfill
\Title{Top quark charge asymmetry measurements with ATLAS detector}
\vfill
\Author{Umberto De Sanctis, \it{on behalf of the ATLAS Collaboration}}
\Address{\Sussex}
\vfill
\begin{Presented}
International Workshop on the CKM Unitarity Triangle (CKM2014)\\
Wien, Austria,  September 8--12, 2014
\end{Presented}
\vfill
\end{titlepage}
\def\thefootnote{\fnsymbol{footnote}}
\setcounter{footnote}{0}

\section{Introduction}

Since its discovery in 1995, the top quark is playing a key role in the understanding of Quantum Chromodynamics (QCD)  processes at high energies. The top quark pair(${t\bar t}$) production cross-section at the LHC  allows to deeply explore the production mechanisms and search for signals of New Physics processes beyond the Standard Model (SM). In this article, the top quark charge asymmetry measurements performed by the ATLAS~\cite{ATLAS_GEN} experiment are presented. Results in single-lepton and dilepton top decay channels for $pp$ collisions at 7 TeV center-of-mass energy using data collected in 2011 are shown.

\section{The top quark charge asymmetry}

At the LHC collider,  ${t\bar t}$ pairs are produced mainly through gluon-gluon ($gg$) fusion process. Only around 20\% of the events are produced from quark-antiquark ($q\bar{q}$) hard collisions, while the fraction coming from quark-gluon ($qg$) partonic processes is almost negligible. The charge asymmetry $A_C$ is a manifestation of the forward-backward asymmetry when the CP invariance holds. It is a tiny NLO QCD effect ($A_C^{SM}$ = 0.0123$\pm$0.0005~\cite{THEOR}) present only in asymmetric initial states, like $q\bar{q}$ and $qg$. In the $t\bar{t}$-system center-of-mass frame, the effect of the charge asymmetry is that tops (antitops) are produced preferentially in the incoming quark (antiquark) direction. At hadron colliders it is difficult to determine the quark/antiquark direction, so another quantity in the laboratory frame is needed to measure this asymmetry. The variable $\Delta y= y_{t} - y_{\bar t}$, where $y$ represents the rapidity of the top/antitop quark, measured in the laboratory frame, is Lorentz invariant. It has the same value as the forward-backward asymmetry in the ${t\bar t}$ center-of-mass frame, computed as a function of the $\cos\theta^*$ angle between the top and the incoming quark. TeVatron experiments used $\Delta y$ variable to measure this asymmetry, counting the number of events where $\Delta y$ is positive or negative.\\
At the LHC, due to the symmetry of the incoming beams, an asymmetry based on the $\Delta y$ variable would vanish. Hence the variable $\Delta |y|= |y_{t}| - |y_{\bar t}|$ has been chosen, based on the fact that quarks are more boosted than antiquarks, due to the different mean momenta carried by valence quarks and sea antiquarks. The asymmetry $A_C$ obtained counting the number of events where $\Delta |y|$  is positive or negative, is called top quark charge asymmetry.

\section{Measurements in the single-lepton channel}
\begin{figure}[!htb]
\begin{center}
\epsfig{file=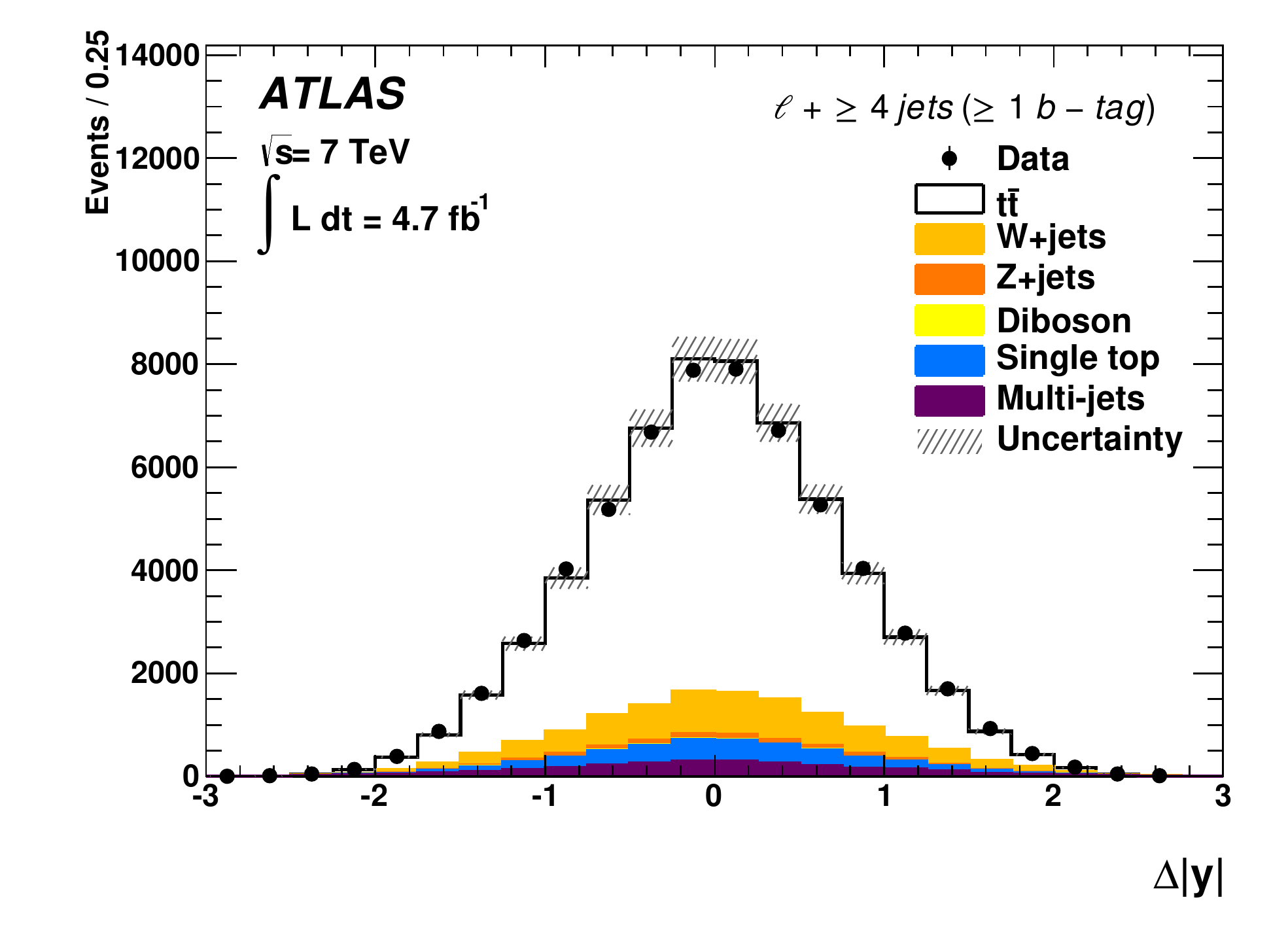,height=1.5in}
\epsfig{file=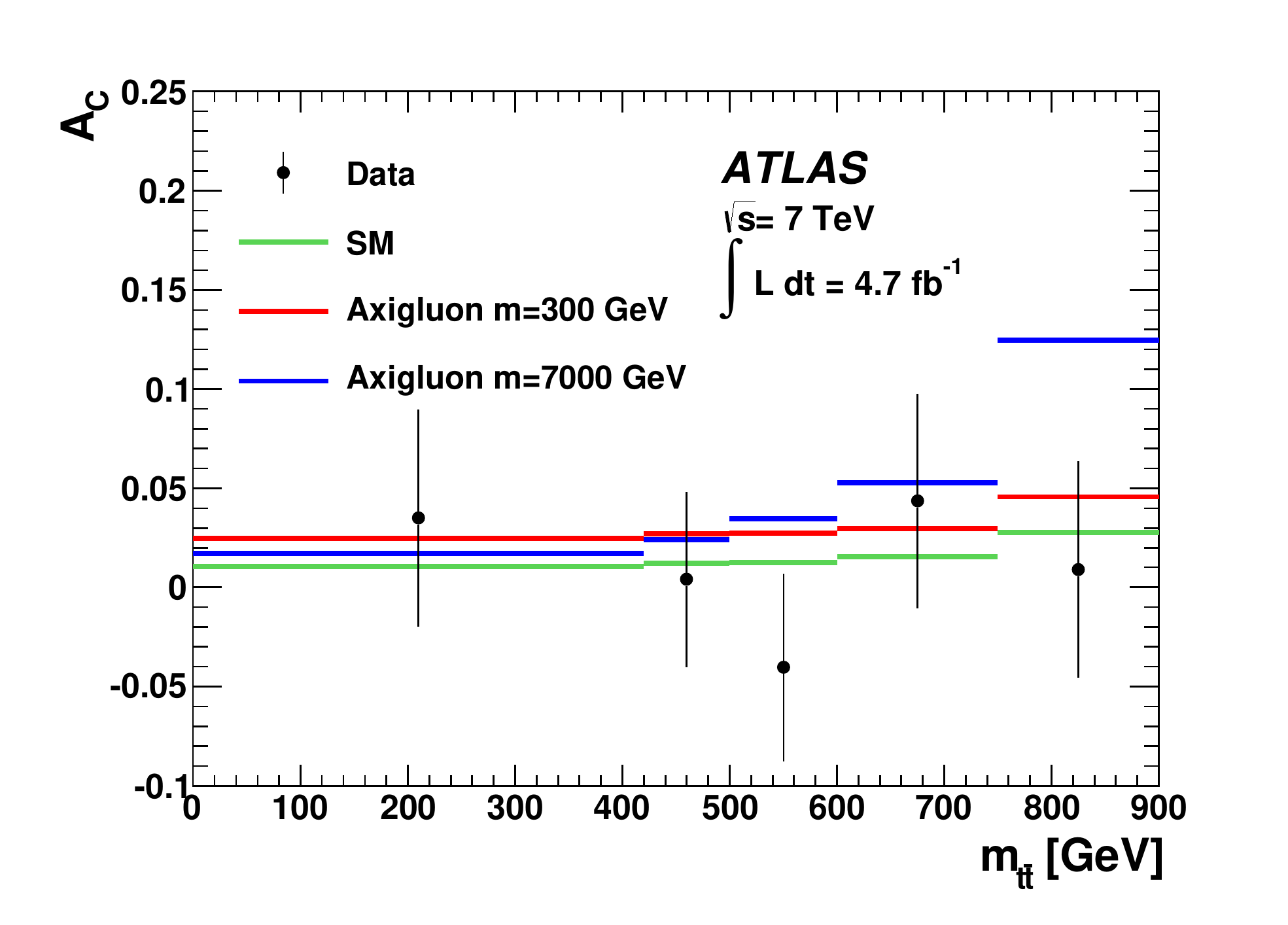,height=1.5in}
\epsfig{file=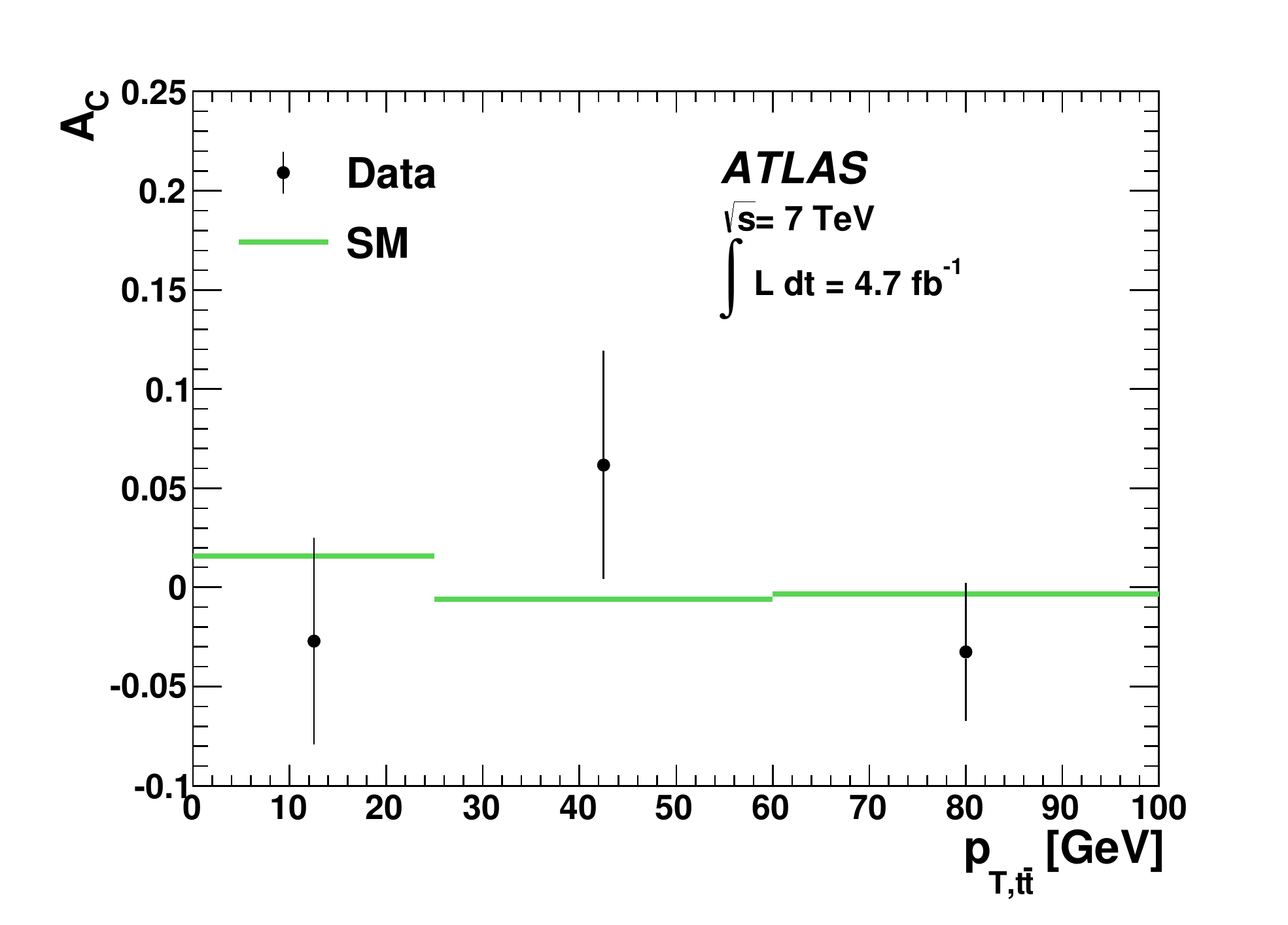,height=1.5in}
\epsfig{file=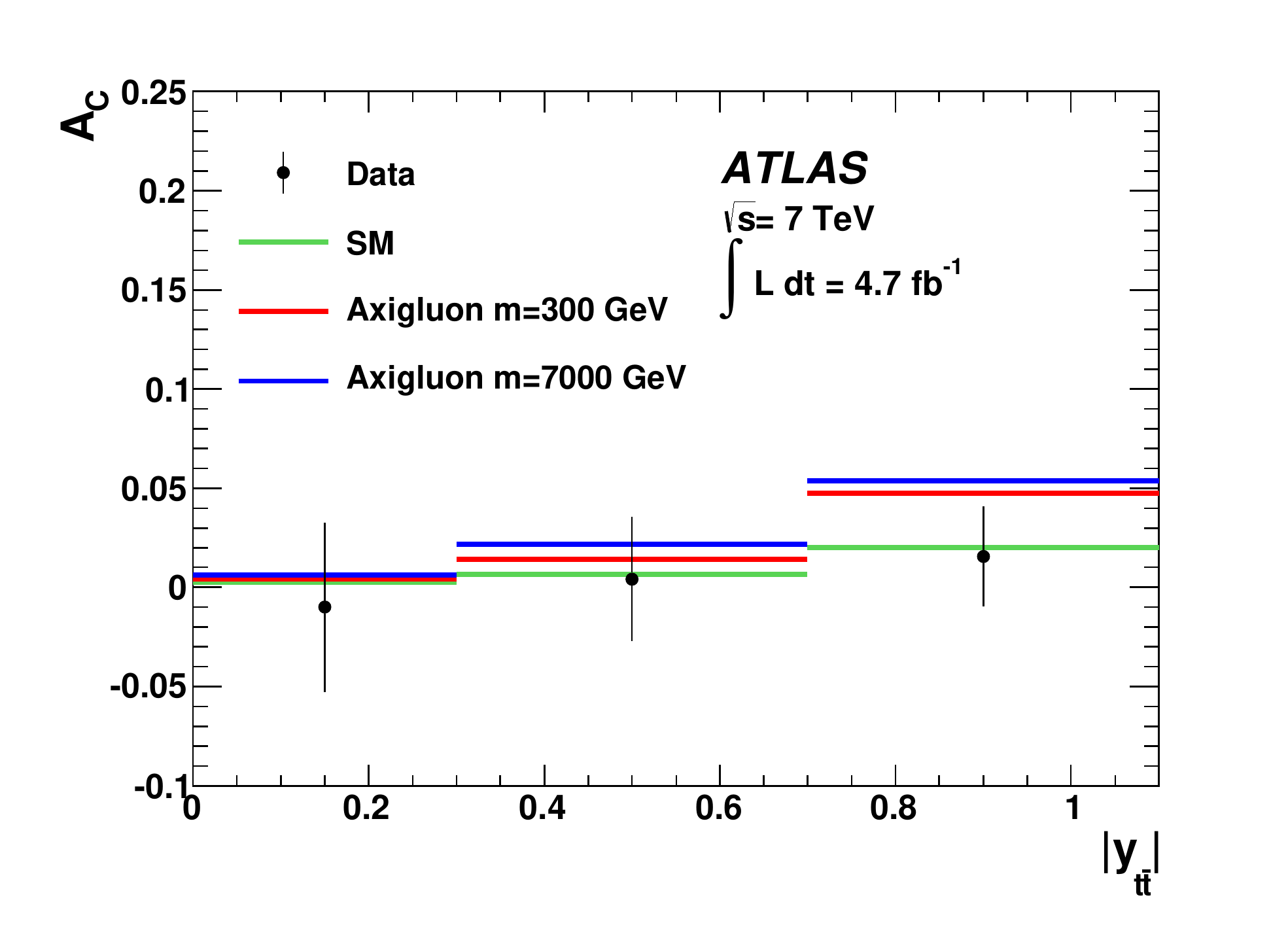,height=1.5in}
\caption{$\Delta |y|$ distribution (top left) and the unfolded $A_C$ asymmetry distributions as a function of $m_{t\bar{t}}$ (top right), $p_{T,{t\bar{t}}}$ (bottom left) and $|y_{t\bar{t}}|$ are shown. The $A_C$ values after the unfolding (points) are compared with the SM predictions (green lines) and the predictions for a colour--octet axigluon with a mass of 300~GeV (red lines) and 7000~GeV (blue lines) respectively. The SM predictions include factorisation and renormalisation scale uncertainties. The values plotted are the average $A_C$ in each bin. The error bars include both the statistical and the systematic uncertainties on $A_C$ values~\cite{ATLAS}.}
\label{fig:atlas1}
\end{center}
\end{figure}

The top quark charge asymmetry $A_C$ has been measured by the ATLAS experiment with data collected at 7 TeV center-of-mass energy corresponding to an integrated luminosity of 4.7 $\rm{fb}^{-1}$~\cite{ATLAS}. The charge asymmetry has been measured inclusively and differentially as a function of the mass ($m_{t\bar{t}}$), the transverse momentum ($p_{T,{t\bar{t}}}$) and the absolute rapidity ($|y_{t\bar{t}}|$) of the $t\bar{t}$-system. In addition, an inclusive and a differential measurement as a function of $m_{t\bar{t}}$ has been  performed with the additional requirement of a minimum velocity $\beta_{z,t\bar{t}} > 0.6$ of the $t\bar{t}$-system velocity along the beam axis to enhance the sensitivity to new physics processes beyond the SM (BSM).\\
Events are selected requiring the presence of exactly one reconstructed isolated electron (muon) with $p_T>25$ (20)~GeV, at least four jets (reconstructed with the anti-$k_T$ algorithm with a 0.4 radius parameter in the $\eta-\phi$ plane) with $p_T>25$~GeV of which at least one tagged as a $b$-jet. Additional cuts are applied on the missing transverse momentum $E_T^{miss} > 30$~GeV and the transverse W mass $m_T^W > 30$~GeV in the electron channel, while a cut on their sum $(E_T^{miss} + m_T^W) > 60$~GeV is applied on the muon channel. In both cases the aim is to reduce the multijet background.\\
The main backgrounds for this analysis, which are multijet and $W$+jets, are estimated using data-driven techniques, while sub-dominant backgrounds like $Z$+jets, single top and diboson ($WW$, $WZ$, $ZZ$) production are estimated using Monte Carlo simulated samples.\\
The $t\bar{t}$ system is reconstructed using a kinematic likelihood fit that assesses the compatibility of the observed events with the topology of a simulated $t\bar{t}$ decays. This method identifies the correct decay topology in 75\% of the cases.\\
The reconstructed $\Delta |y|$ distributions are distorted by acceptance and detector resolution effects. An unfolding procedure is used in order to correct for these effects and to pass from the reconstructed asymmetries to the partonic relative quantities. Simulated  $t\bar{t}$ events are used to build a response matrix $M$ relating true $T$ and reconstructed $R$ observed quantities. Its elements $M_{tr}$ represent the probability and the efficiency of an event produced in the true bin $t$ of the distribution of interest to be reconstructed in any bin $r$. After the subtraction of the backgrounds to the data in the signal region, this matrix is inverted using the FBU (Fully Bayesian Unfolding) technique based on the application of Bayes' theorem to the unfolding problem~\cite{FBU}.\\
Given an observed spectrum $D$ in data for $\Delta |y|$ and the response matrix $ M(T,R)$, the posterior probability density $p$ of the true $T$ spectrum can be computed using the Bayes' theorem:
\begin{equation}
p(T| D,M) \propto L(D| T,M)\times\pi(T)
\end{equation}
where $L(D| T,M)$ is the conditional likelihood for the data $D$ assuming the true spectrum $T$ and the response matrix $M$, and $\pi$ is the prior probability density for $T$.
Assuming that data follows a Poisson distribution, the likelihood $L(D|T,M)$ can be computed using the information of the response matrix $ M(T,R)$ obtained in simulated $t\bar{t}$ events. Conversely, the prior distribution $\pi(T)$ represents our prior knowledge about $T$ before the measurement is performed. In this context, the choice of $\pi(T)$, which is arbitrary, can be interpreted as the choice of a regularisation function in other unfolding techniques. Both a flat prior and a prior distribution based on the curvature of the true $\Delta|y|$ spectrum have been used for the various measurements performed. In both cases it has been checked that any bias in the $A_C$ measurements and their uncertainties was introduced.\\
The inclusive and the differential $A_C$ measurements have been performed combining the electron and muon channels. The inclusive charge asymmetry, together with the measurements and predictions for $m_{t\bar{t}} > 600$ GeV and $\beta_{z,t\bar{t}} > 0.6$ requirements, are shown in Table~\ref{tab:results}.
In Figure~\ref{fig:atlas1} the $\Delta|y|$ distributions after the unfolding procedure and the differential $A_C$ measurements as a function of $m_{t\bar{t}}$, $p_{T,{t\bar{t}}}$ and $|y_{t\bar{t}}|$ are shown.\\
A combination with the inclusive $A_C$ measurement at the same centre-of-mass energy performed by CMS experiment~\cite{CMS} has been also performed. Using the BLUE (Best Unbiased Linear Estimator) method~\cite{BLUE}, the combination is performed taking into account the central values, the statistical and the systematic uncertainties of the two measurements and their correlations. The combined value for the top quark charge asymmetry was found to be: $A_C = 0.005 \pm 0.007(stat.) \pm 0.006(syst.)$, compatible with the SM predictions~\cite{COMB}.

\begin{table}
\begin{center}
\begin{tabular}{|l| c |c|}
\hline
$A_C$ & Data & Theory\\
\hline
 Unfolded & 0.006$\pm$0.010 & 0.0123$\pm$0.0005 \\
 Unfolded with $m_{t\bar{t}} > 600$ GeV & 0.018$\pm$0.022 & $0.0175^{+0.0005}_{-0.0004}$\\
 Unfolded with $\beta_{z,t\bar{t}} > 0.6$ & 0.011$\pm$0.018 & $0.020^{+0.006}_{-0.007}$\\
\hline
\end{tabular}
\caption{\label{tab:results}Measured values of the inclusive charge asymmetry, $A_C$, for the electron and muon channels combined after unfolding without and with the $\beta_{z,t\bar{t}} > 0.6$ cut explained in the text. The $A_C$ measurement with a cut on $m_{t\bar{t}} > 600$~GeV is also shown. SM predictions, as described in the text, are also reported. The quoted uncertainties include statistical and systematic components after the marginalisation~\cite{ATLAS}.}
\end{center}
\end{table}

\section{Measurements in the dileptonic channel}
A measurement of the charge asymmetry has been done by the ATLAS experiment also in the dileptonic $t\bar{t}$ decay channel with an integrated luminosity of 4.7~$\rm{fb}^{-1}$~\cite{ATLAS2}. The events are selected requiring exactly two oppositely charged leptons with the same flavor (i.e. $ee$, $e\mu$ and $\mu\mu$) and $p_T>30(20)$~GeV for electrons (muons). At least two jets with  $p_T>25$~GeV are also required. In the $ee$ and $\mu\mu$ channels additional cuts are applied on the missing transverse momentum $E_T^{miss} > 60$~GeV and the invariant mass of the lepton pair ($m(ll)$) within 10 $GeV$ from the $Z$ boson mass to remove $Z$+jets background. In the $e\mu$ channel a cut on $H_T$, that is the scalar sum of the lepton and jets transverse momentum, is also applied: $H_T>130$~GeV.\\
Having two neutrinos in the final state, the kinematics of the $t\bar{t}$ decays is under-constrained. Hence several combinations of the physical objects in the final state are admissible for each event. In each event, each solution has been weighted according to a likelihood estimator derived from matrix elements for the LO process $gg\to t\bar{t}$. The combination with the highest weight is finally chosen.\\
To measure the asymmetry at the partonic level, a calibration procedure is used in this measurement. After the subtraction of the background, the measured raw asymmetry is calibrated using calibration curves that relate reconstructed and true asymmetries. The curves have been derived from simulated $t\bar{t}$ events where true asymmetries ranging from $-10\%$ to $10\%$ are considered and the corresponding reconstructed asymmetries are evaluated. Raw (after the background subtraction) and calibrated asymmetries in the various channels are shown in Table~\ref{tab:dil}. A combination among the $ee$, $e\mu$ and $\mu\mu$ channels has been then performed taking into account the correlations among the measurements. The combined asymmetry has been measured to be $A_C=0.057 \pm 0.024 (\rm{stat.}) \pm 0.015 (\rm{syst.})$. Furthermore, a combination with a former single-lepton channel measurement performed by ATLAS experiment with 1.04 $fb^{-1}$ of data at a centre-of-mass energy of 7~TeV~\cite{ATLAS3}, has been performed, giving a combined asymmetry $A_C=0.029 \pm 0.018 (\rm{stat.}) \pm 0.014 (\rm{syst.})$.

\begin{table}[!]
\begin{center}
\begin{tabular}{l|cc}  
Channel &  Raw asymmetry & Calibrated asymmetry\\
\hline
$ee$ & $0.051 \pm 0.045 (\rm{stat.})$ & $0.079 \pm 0.087 (\rm{stat.}) \pm 0.028 (\rm{syst.})$ \\
$e\mu$ & $0.037 \pm 0.014 (\rm{stat.})$  & $0.078 \pm 0.029 (\rm{stat.}) \pm 0.017 (\rm{syst.})$ \\
$\mu\mu$ & $-0.001 \pm 0.022 (\rm{stat.})$ & $0.000 \pm 0.046 (\rm{stat.}) \pm 0.021 (\rm{syst.})$\\
\end{tabular}
\caption{Raw and calibrated top charge asymmetries, as desribed in the text, in the three dileptonic channels~\cite{ATLAS2}.}
\label{tab:dil}
\end{center}
\end{table}

\section{Summary}
The top quark charge asymmetry measurements performed by ATLAS experiment at a centre-of-mass energy of 7 TeV have been presented. The asymmetry has been measured in the single-lepton and dileptonic channel depending on the $t\bar{t}$ decay topology. Inclusive and differential measurements, as a function of specific kinematic quantities of the $t\bar{t}$ system has been also presented. A combination with the inclusive $A_C$ measurement obtained by the CMS experiment has been also described. All the presented measurements did not show any significant deviation from the SM predictions.



%
%
%
%
%
 
\end{document}